\documentclass[twocolumn,aps]{revtex4}
\newcommand{\beq}{\begin{eqnarray}}
\newcommand{\eeq}{\end{eqnarray}}
\begin{document}
\title{Active and Sterile Neutrino Emission and SN1987A Pulsar Velocity}
\author{Leonard S Kisslinger, Department of Physics, Carnegie Mellon 
University, Pittsburgh, PA 15213\\ 
Sandip Pakvasa, Department of Physics and Astronomy, University of Hawaii at 
Manoa, Honolulu, HI 96822}

\begin{abstract} 

Recently estimates have been made of the velocities of pulsars produced by 
the emission of sterile neutrinos during the first 10 seconds and by active 
neutrinos during the second 10 seconds after a supernova event reaches 
thermal equilibrium. Neutrinos produced with electrons in the lowest Landau 
level are emitted in the direction of the magnetic field, and the resulting
pulsar velocity depends mainly on the temperature. Using measurements of the
neutrino energies emitted from SN1987A, the temperature can be estimated,
and from this we estimate the velocity of the resulting pulsar from both
active and large mixing-angle sterile neutrinos.

\end{abstract}
\maketitle
\noindent
PACS Indices:97.60.Bw,97.60.Gb,97.60.Jd
\vspace{1mm}

   A supernova event, which is the gravitational collapse of a massive star,
often leads to the formation of a rapidly rotating neutron star, a pulsar.
It has been observed that many pulsars move with linear velocities of 
1000 km/s or greater, the pulsar kick. See  Ref.\cite{hp97} for a review.

   Electrons in very strong magnetic fields, such as those found at the
surface of a protoneutron star created by a supernova, are in Landua levels
\cite{jl,mo}. If the electron is in the lowest Landau level, n=0, it has
only negative helicity with respect to the direction of the magnetic field,
say in the z direction. This leads to neutrinos produced with the electrons
by URCA or modified URCA processes to have momenta strongly correlated 
with the magnetic fields, which could produce pulsar kicks.

  It was shown in a recent work on pular kicks\cite{hjk07} that if electrons 
created by the modified URCA processes,
which dominate neutrino emission after 10 seconds\cite{bw},
are in the n=0 level, only those moving in the z (B) direction will contribute 
to neutrino emision. Therefore, even though only one or two percent of the
neutrino emissivity occurs during the period of approximately 10 to 20 seconds
after the supernova collapse, since almost all emission is correlated
in the z direction this can account for the observed large pulsar velocities,
the pulsar kicks. The resulting pulsar velocity, $ v_{ns}$, is proportional to 
the temperature, T, of the protoneutron star surface to the seventh power,
and one must be able to estimate T in order to predict $ v_{ns}$.

   The largest neutrino emission after the supernova collapse takes place   
during the first 10 seconds, when the neutrinosphere starts at about
40 km, with the URCA process dominating neutrino production. However, due to 
the high opacity for standard model neutrinos in the dense region within the 
neutrinosphere, few neutrinos are emitted, and the large pulsar kick is not
obtained\cite{lq98}. This has led to studies of pulsar kicks coming from
sterile neutrinos to which the active neutrinos oscillate. In a study
using sterile neutrinos with a very small mixing angle and a large mass
(mass $>$ 1 kev), constrained to fit dark matter, it was shown\cite{fkmp03}
that large pulsar kicks can be obtained. More recently, using fits to
the LSND\cite{lsnd} and MiniBoone\cite{mini} experimental data, which seem
to need two large mixing-angle light sterile neutrinos\cite{scs04,ms07,s07,
gsm07}, it was shown\cite{khj09}
that these sterile neutrinos can also give rise to large pulsar velocities.
As in Ref \cite{hjk07} for active neutrinos, it was shown that the pulsar 
velocity from large mixing angle sterile neutrinos is proportional to 
the temperature, T, of the protoneutron star surface to the seventh power.

  In the present work we use the results of Refs \cite{hjk07,khj09} to 
estimate the velocity of a pulsar produced by SN1987A.
In the 10 second period in which the modified URCA process dominates   
neutrino emission, the radius of the neutrino sphere, $R_\nu$, is a little
smaller that the radius of the protoneutron star, $R_{ns}$, so all the created
neutrinos correlated with the z direction are emitted.
In ref\cite{hjk07} it was shown that with the probability of the electron 
being in the n=0 Landau level $\simeq 0.4$ and $R_\nu \simeq 9.96$ km,
the velocity given to the neutron star during this period by active neutrinos,
for a neutron star with the mass of our sun is
\beq
\label{1}
  v_{ns}^{\rm active} &=& 1.03 \times 10^{-4} (\frac{T}{10^{10} K})^7 
\frac{{\rm km}}{{\rm s}} \; . 
\eeq

   During the first 10 seconds, using the model of \cite{ms07,s07} with 
two light large mixing angle neutrinos, it was shown that the velocity
given to a pulsar is
\beq
\label{2}
     v_{ns}^{\rm sterile} &\simeq& 3.35 \times 10^{-7} (\frac{T}{10^{10} K})^7 
\frac{1}{sin^2(2\theta)} \frac{{\rm km}}{{\rm s}} \; ,
\eeq 
where the mixing angles of the two sterile neutrinos give $\sin^2(2\theta)
= 0.004 {\rm \; and \;} 0.2$

   Clearly for a prediction of the velocity of the pulsar, one must know
the temperaure at the surface of the protoneutron star 
quite accurately. If one knows the energy of the emitted neutrinos at 10 
seconds, T is determined by the relationship that $kT=E_\nu/3.15$.

   Twenty neutrinos from SN1987A were detected by Kamiokande-II\cite{hir}
and IMB\cite{bio}. The energies of the neutrinos measured by IMB were two
to three times larger than those of Kamiokande-II. This has been discussed 
in many papers. For the present work we need an analysis of the data to
obtain a mean energy of the neutrinos at about 10 seconds. An early model
\cite{bpps87} chose T=$4.1^{+1.0}_{-0.4}$.  Since then there have been
many analyses.  See references \cite{smir,raff}. Although there
are discrepancies, the general agreement is that the neutrino energy at
10 seconds is in the range 9-14 Mev, giving a temperature range:
\beq
\label{3}
             T &\simeq& (3 \leftrightarrow 4.5) {\rm \;\;MeV}=
(3.5 \leftrightarrow 5.2) \times 10^{10} K \; , 
\eeq
which results in our prediction from Eq.(\ref{1}) that
\beq
\label{4}
     v_{ns}^{\rm active,SN1987A} &\simeq& (0.66 \rightarrow 10.6) 
\frac{{\rm km}}{{\rm s}} \;,
\eeq
which is too small in comparison with other sources of pulsar kicks to be 
significant. Note that if the neutrino energy were 30 MeV, $ v_{ns}$ would
be greater than 1000 km/s,for high-luminoscity pulsars.

   From Eq(\ref{2}) the velocity of the a pulsar produced via SN1987A by
large mixing angle sterile neutrinos has the range given by T and 
$sin^2(2\theta)$
\beq
\label{5}
    v_{ns}^{\rm sterile,SN1987A} &\simeq& (1.08\times 10^{-2} \rightarrow 8.6) 
\frac{{\rm km}}{{\rm s}} \; ,
\eeq
where we use the range $sin^2(2\theta) \simeq 0.004 \rightarrow 0.2$.
Once more, we find that the resulting velocities are too small in comparison 
with other sources of pulsar kicks to be significant. Note also, if the
pulsar momentum were to be produced by the emission of sterile neutrinos,
although one one cannot detect sterile neutrinos and therefore determine the
temperature, if there is a large mixing angle the sterile neutrinos can
oscillate back to active neutrinos, which can be detected.

  In conclusion, we find that the velocity of the pulsar from SN1987A resulting
from the emission of either active or sterile neutrinos is too small to be
significant.

  This research was supported in part by the DOE.

\end{document}